\documentstyle[psfig]{l-aa}

\def\ros{{\sl ROSAT }}
\def\asca{{\sl ASCA }}
\def\G{$\Gamma_{\rm x}$ }
\def\approxlt{\mathrel{\hbox{\rlap{\lower.55ex \hbox {$\sim$}}
        \kern-.3em \raise.4ex \hbox{$<$}}}}
\def\approxgt{\mathrel{\hbox{\rlap{\lower.55ex \hbox {$\sim$}}
        \kern-.3em \raise.4ex \hbox{$>$}}}}

\begin{document}

  \thesaurus{03         
               (11.01.2;  
               11.09.1;  
               11.19.3;  
               13.25.2)  
}

   \title{The \ros view of NGC\,1365 and the luminous highly variable 
off-nuclear
X-ray source NGC\,1365--X1} 
   \author{ Stefanie Komossa\inst{1} \and Hartmut Schulz\inst{2}}
\offprints{Stefanie Komossa,\\
 skomossa@xray.mpe.mpg.de}
\institute{
Max-Planck-Institut f\"ur extraterrestrische Physik,
         Postfach 1603, D-85740 Garching, Germany
\and
Astronomisches Institut, Ruhr-Universit\"at,
              D-44780 Bochum, Germany}
\date{Received: 28 May 1998; accepted: August 1998}
   \maketitle
\markboth{St.~Komossa, H.~Schulz: The \ros view of NGC\,1365 and 
          NGC\,1365-X1}
{St.~Komossa, H.~Schulz: The \ros view of NGC\,1365 and NGC\,1365-X1}

   \begin{abstract}

We present an X-ray spectral and spatial analysis of the
composite starburst/Seyfert galaxy NGC\,1365.
Excellent fits of the \ros PSPC spectrum
are obtained by combining a soft thermal component with a hard
powerlaw. 
The hard component may be either seen directly
or can be explained by 
scattering of the AGN powerlaw at circumnuclear warm high-column-density 
gas. 
A compilation of the multi-wavelength properties of NGC\,1365
and comparison with hard X-ray
selected AGNs shows that the hard component of NGC\,1365 is too
faint compared to its broad Balmer line components challenging
simple unified models.
According to analytical estimates, supernova driven outflow can fully
account for the X-ray luminosity in the Raymond-Smith component
if the observed IR emission is mainly provided by the central starburst.  
We do not find obvious optical counterparts for
three faint PSPC sources south of the nucleus. In particular, there is 
no coincidence with the two supernovae reported in NGC\,1365. 

With the \ros HRI data, we have precisely located the extraordinary 
southwest X-ray source  
NGC\,1365--X1 which
falls on one of the subordinate 
spiral arms. The source is found to be highly variable
(a factor $\approxgt$ 10) on the timescale of months.
Intrinsic to NGC\,1365, its huge luminosity makes it exceptional among stellar
X-ray sources. At present, the most likely interpretation seems to be 
an ultra-powerful X-ray binary. 

   \end{abstract}
%
\section{Introduction}
NGC\,1365 is a prominent  
barred spiral in the southern hemisphere.
Its nuclear and disk emission-line gas has been
investigated in numerous optical studies (e.g., Burbidge \& Burbidge 1960,
Osmer et al. 1974, 
Veron et al. 1980, Alloin et al. 1981,  
Edmunds \& Pagel 1982, Phillips et al. 1983, J\"ors\"ater
et al. 1984a,b,
Teuben et al. 1986,
Edmunds et al. 1988, Schulz et al. 1994, Roy \& Walsh 1988, 1997).
The presence of an AGN was first suggested by Veron et al.\ (1980) who 
found  broad emission-line H$\alpha$ indicative of a Seyfert-1.5 galaxy.
The FWHM of the broad component is 1900 km/s (e.g., Schmitz 1996).   
Surprisingly, just in the nucleus identified
by Edmunds \& Pagel (1982), Seyfert-typical narrow-line emission
line ratios are missing, probably due to an interference with HII 
regions. 
To the SE of the nucleus, [OIII]$\lambda5007$ 
enhancements and occasional line splitting apparently related to the AGN
are nevertheless present. 
They trace a wide cone reminiscent of the supposed radiation cones in
Seyfert-2 objects (J{\"o}rs\"ater \& Lindblad 1989, 
Storchi-Bergmann \& Bonatto 1991, Kristen et al.\ 1997).
Outflow cone models were employed to fit the kinematics in this
region (Phillips et al.\ 1983, Edmunds et al.\ 1988,  Hjelm \& Lindblad 
1996).

In the remaining circumnuclear emission-line region, HII region-like line 
ratios
are common indicating widespread circumnuclear star formation.
Two outstanding hot spots 7\arcsec~ SW of the center corroborate the 
starburst character.
Sandqvist et al.\ (1995) mapped abundant molecular gas in the center
which may provide the fuel for
the birth of stars and which could have developed the molecular torus
believed to be a prerequisite of Seyfert unification (Antonucci 1993).  
  
Summarizing, optical studies suggest that the central region
of NGC\,1365 consists of an AGN of apparent low luminosity
surrounded by a region of enhanced
star formation.  
However, the
relationship between the stellar and nonthermal activity
and the geometry of the nucleus need further scrutiny.
 
X-rays are an important probe of the central activity.
In AGNs, they are believed to arise in the innermost core.
Interestingly, NGC\,1365 lies
deep within the {\em narrow-line} galaxies in a plot of
$L_{\rm X}$ versus $L_{60\mu m}$ (Green et al. 1992)
although its Balmer lines have a conspicuous {\em broad-line} component
signifying a class-1.5 Seyfert.
In an early study of \ros PSPC X-ray spectra by Turner et al. (1993; TUM93 hereafter)
a powerlaw fit was found to be unsatisfactory unless a line at 0.8 keV 
or a Raymond-Smith component was added.   
Recent detection of conspicuous FeK line emission in \asca data
by Iyomoto et al.\ (1997; I97 hereafter) supports the view that the hard X-rays 
trace genuine 
nonthermal activity in NGC\,1365. In particular, the \asca spectra
revealed a striking similarity to those of NGC\,1068, the prototype
of a hidden class-1 Seyfert (Antonucci \& Miller 1985).

In the present work, we perform a detailed analysis of all  
\ros X-ray observations of the core of NGC\,1365 and an extraordinary 
off-nuclear X-ray source which we term NGC\,1365--X1. 
This investigation includes new PSPC data and spectral models 
like a `warm reflector', 
and the first high-spatial resolution X-ray study based on 
HRI observations retrieved from the archive.

Basic data of NGC\,1365 are adopted from the compilation in Tab. 2 of
Schulz et al.\ (1994): $v_{\rm sys} = (1639\pm20)$ km/s implying a 
distance of
19.8 Mpc (linear scale 96 pc/\arcsec) with $H_0 = 75$ km/s/Mpc  
and
the virgocentric model of Kraan-Korteweg (1986).
Recent HST borne Cepheid data lead to an insignificantly smaller distance
of $18.4 \pm 1.8$ Mpc (Madore et al.\ 1996, Silbermann et al. 1998). 
J{\"o}rs{\"a}ter \& van 
Moorsel (1995)
found $v_{\rm sys} = 1632$ km/s and suggest a moderate revision of
geometric data (PA of line of nodes and value of inclination).

The paper is organized as follows: In Sect.\,2 the X-ray spectrum of the 
nuclear
source is analyzed while in Sect.\,3 the high-spatial-resolution HRI data
revealing the core source and the variable enigmatic source NGC\,1365--X1 
are
presented. 
In Sect.\,4, the properties of NGC\,1365--X1 are investigated in more detail.  
The nature of the sources is discussed in Sect.\,5 which is 
followed
by the concluding summary in Sect.\,6.  
    
\section{Data analysis -- PSPC}
\subsection{Data reduction} 
NGC\,1365 is serendipituously located in the field of view of a ROSAT
(Tr\"umper 1983)  PSPC (Pfeffermann et al.\ 1987) observation
performed from Jan. 30 -- Feb. 2, 1993. 
The total exposure time is 7.4 ksec. The source is located at an off-axis angle
of 30\arcmin.

For further analysis, the source photons were extracted
within a circular cell of radius 4\arcmin~ centered on the core of NGC\,1365. 
The background was determined in a source-free region near
the target source and subtracted. 
The data were corrected for vignetting 
and dead-time, using the EXSAS software package (Zimmermann et al.\ 1994a).
The mean source countrate is $0.10\pm{0.01}$ cts/s.

To carry out the spectral analysis source photons in
the amplitude channels 11-240 were binned
according to a constant signal/noise ratio of 5$\sigma$.

   \begin{table}            
     \caption{Log of observations. $t_{\rm exp}$ gives the exposure time in ksec,
             $CR_{\rm c}$ and $CR_{\rm X1}$ refer to the countrate in cts/s of the core source and 
          the off-nuclear source NGC\,1365-X1, respectively.}
     \label{obslog}
      \begin{tabular}{lllcl}
      \hline
      \noalign{\smallskip}
        date & $t_{\rm exp}$ & $CR_{\rm c}$ & $CR_{\rm X1}$ & obs. label \\
       \noalign{\smallskip}
      \hline
      \noalign{\smallskip}
 ~~PSPC \\ 
 Jan 30 - Feb 2, 1993 & 7.4 & 0.102 & --$^{*}$ & PSPC-1 \\
 Feb 5 - Feb 10       & 7.7 & 0.096 & 0.005 & PSPC-2 \\
\noalign{\smallskip}
 ~~HRI \\
 Jul 20 - Aug 4, 1994 & 9.8 & 0.020 & 0.0026 & HRI-1 \\
 Jul 4 - Jul 5, 1995 & 9.8 & 0.018 & $<$0.001 & HRI-2 \\
      \noalign{\smallskip}
      \hline
      \noalign{\smallskip}
  \end{tabular}

\noindent{\footnotesize $^{*}$ not resolved well enough in this off-axis observation;
thus, no countrate for NGC\,1365-X1 was determined }
   \end{table}

\subsection{Spectral analysis}
A single powerlaw (PL) provides a marginally successful fit to the X-ray 
spectrum ($\chi{^{2}}_{\rm red}$ = 1.2)
with a photon index $\Gamma_{\rm x} = -2.7$, 
a 1-keV normalization of $4.4\,10^{-4}$ ph/cm$^2$/s/keV,
and cold absorption of $N_{\rm H} = (0.8\pm{0.4})\,10^{21}$ cm$^{-2}$
which is larger than the Galactic value 
in the direction of NGC\,1365, $N_{\rm Gal} = 0.135\,10^{21}$ cm$^{-2}$
(Dickey \& Lockman 1990).
The fit becomes worse if we fix the cold
absorption to the Galactic value ($\chi{^{2}}_{\rm red}$ = 3.1).
The PL fit leaves systematic residuals around 0.6--1 keV.
This new observation confirms the results of TUM93, 
who analyzed a different PSPC
observation and tentatively invoked a line of high equivalent width
at 0.8 keV.  (Re-analyzing this latter data set, using the same 
data reduction techniques as in the observation presented above 
and the same powerlaw model, we find both spectra to be consistent 
with each other.)

   \begin{table}
     \caption{Summary of spectral fits to the core emission of NGC\,1365 
(PL = power law,
RS = Raymond-Smith model of cosmic abundances).
$T$ = temperature of
RS component, $N_{\rm H}$ = column density of cold absorber.
For the model of a warm reflector, see text.
Instead of individual error bars, we list different models that
successfully describe the data.
              }
     \label{fitres}
      \begin{tabular}{lllccc}
      \hline
      \noalign{\smallskip}
        model & $N_{\rm H}$ & $\Gamma_{\rm x}$ & Norm$_{\rm pl}$ & $kT$
                            & $\chi^2_{\rm red}$ \\
       \noalign{\smallskip}
      \hline
       \noalign{\smallskip}
              & 10$^{21}$ cm$^{-2}$ & & $^{(3)}$ & keV &  \\
       \noalign{\smallskip}
      \hline
      \hline
      \noalign{\smallskip}
PL & 0.8 & $-2.7$ & 4.4\,10$^{-4}$ & -- &
1.2  \\
RS & 0.135$^{1}$ & -- & -- & 1.2 &
1.5 \\
RS$^{2}$ & 0.22 & -- & -- & 1.1 &
0.9 \\
\hline
\noalign{\smallskip}
RS+PL-1.9 & 0.135$^{1}$ & $-1.9^{1}$ & 9.9\,10$^{-5}$ & 0.95 &
0.9 \\
RS+PL-1.0 & 0.135$^{1}$ & $-1.0^{1}$ & 1.7\,10$^{-4}$ & 0.72 &
 0.5 \\
RS+RS & 0.135$^{1}$ & -- & -- & 0.7/6.3
& 0.5  \\
      \noalign{\smallskip}
      \hline
      \noalign{\smallskip}
  \end{tabular}

\noindent{\small $^{(1)}$ fixed;
$^{(2)}$ abundances of 0.2\,$\times$\,cosmic; $^{(3)}$ PL normalization
at 1 keV in ph/cm$^2$/s/keV \\
}
  \vspace{-0.4cm}
   \end{table}

We then applied models involving emission from a Raymond-Smith (RS)
plasma. If not noted otherwise, cosmic abundances (Allen 1973) were assumed
throughout as further justified in Sect. 5.
A one-component RS model yields an inferior fit,
with $\chi{^{2}}_{\rm red}$ = 1.5.
The quality of the fit can be improved by allowing for low  
metal abundances. Fixing them to 0.2 $\times$ cosmic
leads to $kT$ = 1.1 keV, $N_{\rm H}$ = $0.22\,10^{21}$ cm$^{-2}$,
 and $\chi{^{2}}_{\rm red}$ = 0.9.

\begin{figure}[t]
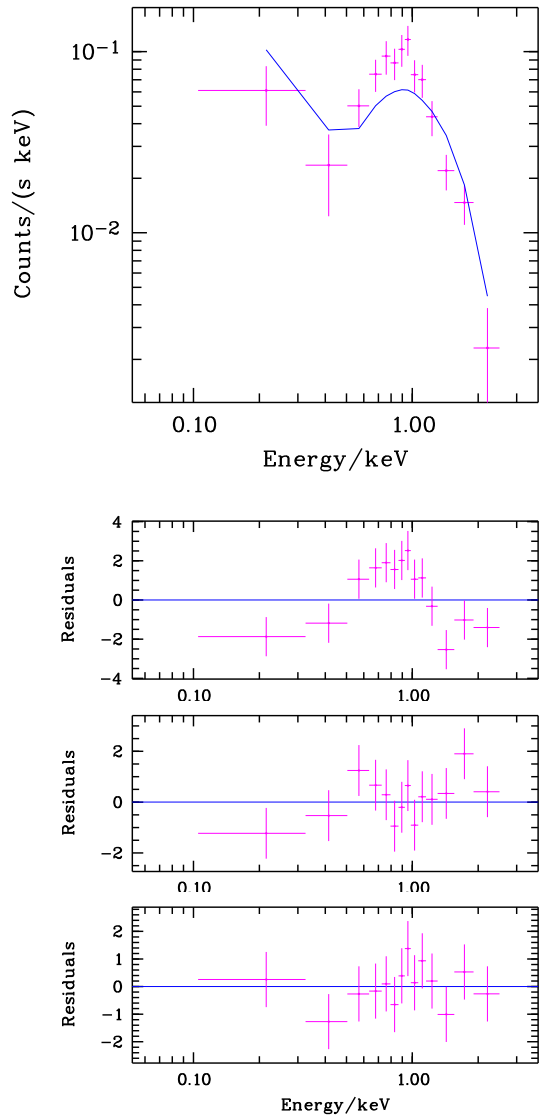

      \vbox{\psfig{figure=H1039.F1a,width=7.3cm,%
          bbllx=2.5cm,bblly=1.1cm,bburx=10.1cm,bbury=11.7cm,clip=}}\par
            \vspace{-0.7cm}
      \vbox{\psfig{figure=H1039.F1b,width=7.3cm,%
          bbllx=2.5cm,bblly=1.1cm,bburx=10.1cm,bbury=4.5cm,clip=}}\par
            \vspace{-0.7cm}
      \vbox{\psfig{figure=H1039.F1c,width=7.3cm,%
          bbllx=2.5cm,bblly=1.1cm,bburx=10.1cm,bbury=4.5cm,clip=}}\par
            \vspace{-0.4cm}
\caption[SEDx]{The first panel shows the observed X-ray spectrum of 
NGC\,1365 (crosses)
and the pl model fit. The second panel displays the deviation between
data and this model (residuals) in units of $\sigma$. 
For comparison, the third panel gives the residuals for the RS+PL
fit. 
The lowest panel displays the residuals for the warm reflector model. 
  }
\label{SEDx}
\end{figure}

In the next step, two-component models were compared with the data. 
Both, a two-temperature RS model, and a RS+PL model provide
excellent descriptions of the data.
For the RS+RS model (and $N_{\rm H}$ fixed to the Galactic value), we 
derive
temperatures of 0.7 keV and 6.3 keV, with the hotter component dominating
in flux ($\chi{^{2}}_{\rm red}$ = 0.5).
For the RS+PL description ($N_{\rm H}$ and photon index $\Gamma_{\rm 
x}=-1.9$
fixed) we find a temperature of 0.95 keV ($\chi{^{2}}_{\rm red}$ = 0.9).
Treating $N_{\rm H}$ as additional free parameter results 
in a value slightly larger than the Galactic value, but consistent with
the latter within the errors. The residuals for this model
are shown in Fig. \ref{SEDx}. 
If we invoke a flatter underlying powerlaw spectrum,
the temperature of the RS component slightly decreases. E.g., for
$\Gamma_{\rm x}=-1.0$ (about the value found in one of the fits 
in I97) and $N_{\rm Gal}$ we get $kT = 0.72$~keV and
$\chi_{\rm red}^2 = 0.5$.
The temperature rises somewhat, and the relative contribution of the RS
component increases, if we allow for non-cosmic abundances
(e.g., $T$ = 1.0 keV and $\chi{^{2}}_{\rm red}$ = 0.9 for metal
abundances of 0.3 $\times$ cosmic).
   
Finally, we checked whether the data can be described in terms of 
a lowly-ionized warm absorber, or a dusty warm absorber,
or a warm reflector
(the models were calculated with Ferland's (1993) code {\em Cloudy};
for assumptions and a detailed description see Komossa \& Fink 1997a,b). 
Warm absorbers of low ionization or dusty ones
imprint spectral signatures mainly at low energies 
beyond the \asca sensitivity range (e.g., a carbon absorption edge
at $\sim$0.3 keV). 
The 
underlying powerlaw index was fixed to the canonical Seyfert value
of $-1.9$; solar abundances were adopted. 
Whereas a warm absorber seen purely in absorption does not give a 
successful
spectral fit, one in reflection provides an excellent description of the 
data. The column density of the warm material is of order 
$N_{\rm w} \simeq 10^{23}$ cm$^{-2}$ and the 
ionization parameter{\footnote{ the ionization parameter is defined as
$U=Q/(4\pi{r}^{2}n_{\rm H}c)$, where $Q$ is the
number rate of photons above 13.6 eV, $r$ is the distance
nucleus -- warm absorber and $n_{\rm H}$ the hydrogen density}}   
$U \simeq 10$ ($\chi{^{2}}_{\rm red}$ = 0.6; the fit residuals are 
displayed in Fig. \ref{SEDx}). The model is very similar to the
scattering model that we applied to the X-ray spectrum of NGC\,6240
in Komossa et al. (1998).

\subsection {Temporal analysis}

The PSPC X-ray lightcurve of NGC\,1365 was derived. The X-ray 
emission is found to be constant.  
In the \ros all-sky survey, NGC\,1365 was detected with a countrate 
of 0.09$\pm{0.03}$ cts/s, within the errors consistent with the value of 
the pointed observations carried out later.  

\subsection {Spatial analysis}

\begin{figure}[t]      
      \vbox{\psfig{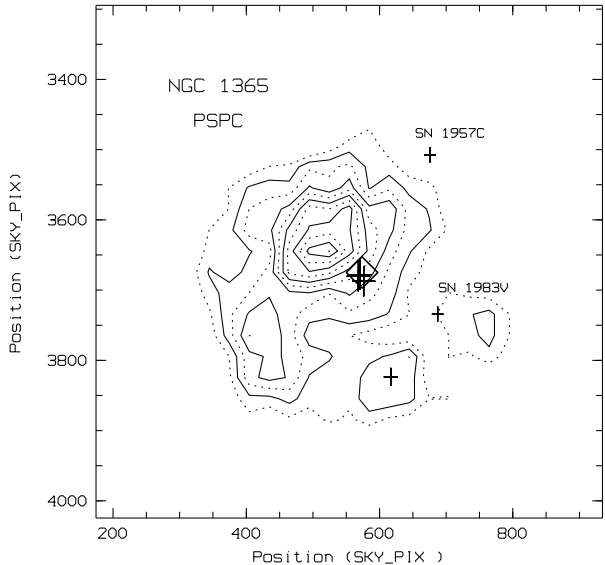}}\par
\caption[chi]{
Contour plot of the X-ray emission from the direction of NGC
1365. The symbols mark the locations
of the core emission in the other observations
and the positions of further sources.
Explicitly, the lozenge gives the position of the optical nucleus of
NGC\,1365, and the crosses the positions of the X-ray maxima for the
two HRI observations and the other PSPC observation. The coordinates
agree very well. The offset of the X-ray maximum in the present PSPC 
observation
can be traced back to the known boresight error of the telescope.
Also marked (large cross) is the HRI position of
NGC\,1365--X1 and the optical locations of two supernovae observed in NGC 1365.
Contour levels range from 2 to 7$\sigma$ above the background and are
plotted in steps of 0.5$\sigma$. Every second contour is plotted as dotted line.
2 sky-pix correspond to a scale of 1\arcsec.
}
\label{chi}
\end{figure}

Fig. \ref{chi} shows the PSPC X-ray contours for the emission from the
direction of NGC\,1365. Four sources are detected, three of them very 
weak. 
None of them coincides with any of the two supernovae 
reported in NGC\,1365, SN 1957C and SN 1983V, but we note that `companion 4' of
TUM93 (their Tab. 2A, 2B) is at 16\arcsec$-$20\arcsec~ distance to SN 1957C
(cf. Fig. 3 of Komossa \& Schulz 1998).
Several source positions are marked in Fig. 2, including
that of the enigmatic source NGC\,1365--X1.   

Whereas the X-ray emission maximum of the HRI observations
and a second PSPC observation (see below) coincide well with the 
optical position of the nucleus, the maximum in the present 
PSPC observation shows a slight off-set. This can be traced back to
the known boresight error of the telescope (e.g. Briel et al. 1994).  

\section{Data analysis -- HRI}

\subsection{Data reduction} 

The region around NGC\,1365 was observed twice with the HRI,
from July 20 -- August 4, 1994 (labeled HRI-1 hereafter) and July 4 -- 5, 
1995
(HRI-2) with an
exposure time of 9.8 ksec each.
Both observations are centered on NGC\,1365.
Nine X-ray sources are detected with a likelihood $>10$
within the field of view in the first observation,
eight sources in the second.
Here, we focus on emission from the core
of NGC\,1365 and the off-nuclear bright source RX\,J0333 --3609,
hereafter referred to as NGC\,1365--X1.
Determining the background in source-free regions
near the target sources, we find background-corrected mean source 
countrates
of 0.020$\pm{0.002}$ cts/s and 0.018$\pm{0.001}$ cts/s
for NGC\,1365 for HRI-1 and HRI-2, respectively.
NGC\,1365--X1 is only detected in the first observation, with a countrate 
of
0.0026$\pm{0.0006}$ cts/s. In the second observation, we estimate an
upper limit on the countrate of $CR$ $<$ 0.001 cts/s.

We also re-analyzed another PSPC observation of 7.7 ksec duration
(previously presented in
TUM93; these data are referred to as `PSPC-2' hereafter) 
to check for the presence and brightness
of NGC\,1365--X1. We clearly detect this source (note that it is  
none of the sources reported in TUM93; their Tab. 2B) 
and derive a countrate of
$\sim$ 0.005 cts/s.

\subsection {Spatial analysis}

In the HRI observation the two strongest central 
sources are clearly separated for the 
first time.
The stronger one coincides well with the nucleus of NGC\,1365,
the other source, NGC\,1365 -- X1, lies to the south of the bar within or
projected onto  
one of the `side' spiral arms (Fig. \ref{over}). 

  \begin{figure}[t]      
      \vbox{\psfig{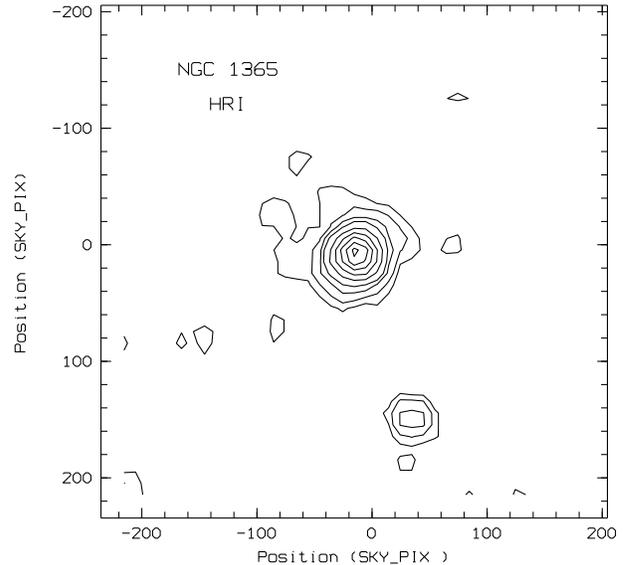}}\par
 \caption[conthri]{HRI contour plot of the X-ray emission from the
direction of NGC\,1365. The lowest contour is at 1$\sigma$ above 
the background, the succeeding ones are at 1.5, 2.5, 3.5$\sigma$ and then
increase to 13.5$\sigma$ in steps of 2$\sigma$. 
The X-ray maximum coincides with the optical nucleus. 
The source to the SW is NGC\,1365--X1. 
}
 \label{conthri}
\end{figure}
%
\begin{figure*} 
\vspace{0.4cm}
\vbox{\psfig{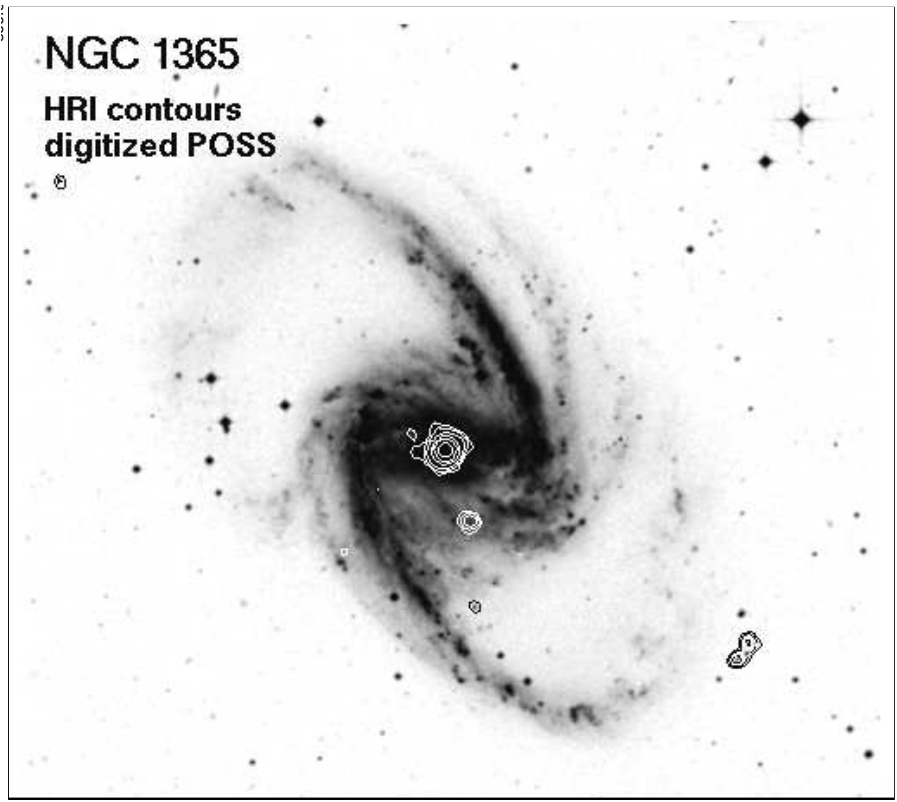}}\par   
\hfill
\begin{minipage}[]{0.27\hsize}\vspace{-3.0cm}
\hfill
\caption[over]{Overlay of the HRI X-ray contours on an optical image
of NGC\,1365 from the digitized POSS.
}
\label{over}
\end{minipage}
\end{figure*} 
The X-ray positions of the nuclear source (J\,2000) are
$\alpha = 3^h 33^m 36\fs4,\delta = -36\degr 8\arcmin 26\farcs7$ (HRI-1)
and $\alpha = 3^h 33^m 36\fs6, \delta = -36\degr 8\arcmin 28\farcs1$ 
(HRI-2),
which compares to the position of the optical nucleus at
$\alpha = 3^h 33^m 36\fs4, \delta = -36\degr 8\arcmin 25\farcs5$ taken 
from NED.
The coordinates of the X-ray emission maximum of NGC\,1365--X1 are
$\alpha = 3^h 33^m 34\fs5, \delta = -36\degr 9\arcmin 38\farcs0$ (HRI-1).

A contour plot of the X-ray emission traced by the HRI
is shown in Fig. \ref{conthri} (see also Fig. \ref{over}). 
Due to its higher spatial resolution the HRI contours reveal
well separated sources. Because of the HRI's lower sensitivity,   
some of the faint extranuclear PSPC sources are undetected.    
`Companion 2' of TUM93 is clearly resolved in two sources
(Fig. \ref{over}). 
Although the core source seems to be extended on weak emission levels,
the presumably pointlike BL Lac object in the field of view shows similar
deviations from the instrumental PSF. A deeper observation with, e.g., AXAF 
would be needed 
to assess if part of the core emission is extended. 

We note in passing that NGC\,1365--X1 lies on a line connecting 
the nuclear source and the X-ray bright BL Lac object MS\,03313-36;
several further sources are `aligned' relative to the core source  
(cf. Fig. 3 of Komossa \& Schulz 1998; for a discussion of aligned X-ray 
sources in/around nearby galaxies
see e.g. Arp 1997,
Radecke 1997).

\subsection {Temporal analysis}

We produced long- and short-term X-ray lightcurves for both sources.
Whereas the emission from the core of NGC\,1365 and NGC\,1365--X1
is constant within the errors on short timescales, 
that of NGC\,1365--X1 is strongly
variable on a longer timescale.
A clear detection during the first HRI observation in 1994 was followed
by a non-detection one year later implying a countrate drop  
by more than a factor 2. 
For a comparison of the HRI counts with the PSPC and \asca data, we 
converted the counts from
the latter instruments to equivalent
\ros HRI counts, assuming a spectral shape as estimated in 
I97 ($\Gamma_{\rm x} = -1.7$ and their Galactic 
$N_{\rm H}$ = 0.15\,10$^{21}$ cm$^{-2}$; see also our Sect. 4).
The following countrates are obtained: $CR_{\rm HRI}$ = 0.0014 (PSPC-2; 
2/1993),
0.0026 (HRI-1; 7/94), 0.0095 ({\sl ASCA}; 1/95), $<$ 0.001 cts/s 
(HRI-2; 7/95).

\begin{figure}[t]
      \vbox{\psfig{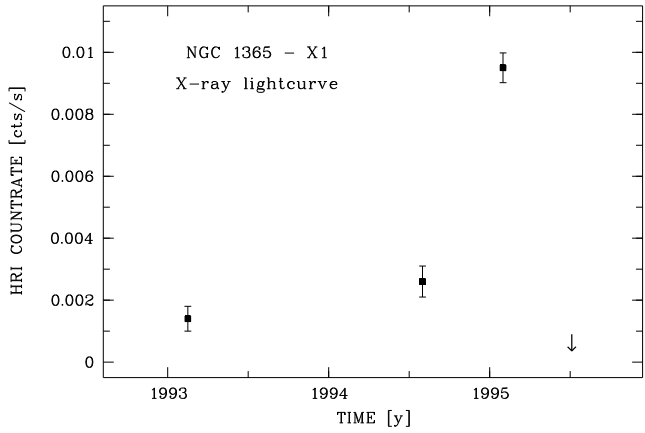}}\par
 \caption[light_offn]{Long-term X-ray lightcurve of NGC\,1365--X1, 
produced  as
described in the text. The source is strongly variable on the timescale
of months.  
}
 \label{light_offn}
\end{figure}

So far, the highest known state, corresponding 
to an isotropic luminosity of $L_{\rm 2-10 keV} \simeq 4\,10^{40}$ 
erg/s at the distance of 
NGC\,1365,
was reached during the \asca observation (I97). 
Half a year later, during the
HRI-2 observation, the source had dropped by a factor $\ge$ 10 as compared 
to the `high state'.
The X-ray lightcurve is displayed in Fig. \ref{light_offn}.  
%

\section{NGC\,1365--X1: Safety checks}

Given the extraordinary properties of NGC\,1365--X1, we made several
checks to test the robustness of the results (concerning amplitude
of variability, derived luminosity, assumed spectrum):

(i) First, we note, that measurements with the {\em same} instrument
under similar conditions exhibit clear
variability of the source: It is only detected in
one out of the two HRI observations which have the same exposure
time and pointing direction. NGC\,1365--X1 is immediately visible 
in the first exposure, but absent in the second.
A similar argument holds for the two ASCA observations of 
I97.

(ii) The total {\em amplitude} of variability depends on the
inter-instrument countrate conversion (from \asca-SIS to \ros-HRI).
To test the reliability of the conversion, we also converted
the \asca countrate of the central source (which did not show 
variability
during the \ros or the \asca observations) to HRI countrate
and we find good agreement. (Explicitly, starting with CR = 0.023 cts/s 
(\asca SIS)
and assuming constant spectral shape (we used the RS model),
we derive ${CR}_{\rm HRI}$ = 0.017$\pm{0.001}$, in agreement with the 
observed value of
0.019$\pm{0.002}$.)

(iii) The {\em luminosity} of the source depends on the assumed
spectral shape and is particularly sensitive to the value of
cold absorption.
I97 used the Galactic $N_{\rm H}$ only, which  
corresponds
to a lower limit on source luminosity.
To get a crude spectral shape in the \ros band,
we used the on-axis PSPC observation (PSPC-2) and extracted
source photons in a circle around NGC\,1365--X1.
A potential problem is the closeness to the nuclear source,
and the weakness of the target source.
Clearly aware of being at the limit of the applicability
of the chi-square fit procedure, we fit a powerlaw
to the source spectrum. In a first step, we fixed $N_{\rm H}$ to the
Galactic value. We derived a photon index of $\Gamma_{\rm x} = 
-1.5\pm{0.5}$
which is in good agreement with the \asca value of $-1.7$.
We then repeated the spectral fit for a second background region;
the results are the same within the errors.
This insensitivity to the background chosen also implies that 
a potential contribution to the X-ray spectrum from other spiral arm
sources in NGC\,1365 or weak extended emission from the Fornax cluster
is negligible. 

We note that the observed (isotropic) high-state luminosity is a {\em 
lower}  
limit to the total luminosity (except if dominated by beaming) 
since the source may not have been observed exactly at maximum 
light, and the X-ray spectrum may extend into the EUV and/or have a soft
excess added. 

(iv) Finally, we recall that a foreground or 
background object can not be ruled out at present.  
We consider this unlikely, though, due to the following reasons:
(a) bright off-nuclear X-ray sources are being observed in more
and more nearby galaxies (the present case just seems to be an extreme 
one);
(b) foreground object (e.g.\ a Galactic X-ray binary): 
Given the high Galactic lattitude of NGC\,1365  
($-54\fdg6$)
a chance coincidence has a low probability; 
(c) background quasar: located behind the dusty/gaseous spiral arm 
seen at
the position of NGC\,1365--X1, the cold absorbing column in the direction 
of a background object
should significantly exceed the Galactic value. However, 
the source is
separately detected in the soft (below 1 keV) and the hard band.  
We again used the crude PSPC spectrum described above, 
fixed \G = --1.9 and treated $N_{\rm H}$ as free
parameter for an order-of-magnitude estimate of its value.  
This yields $N_{\rm H} = (0.21\pm0.15)\,10^{21}$~cm$^{-2}$ and is 
consistent
with the Galactic value or only slightly larger.
Again, we repeated the spectral fit for the second background region
and get the same results within the errors.

\begin{table}             
\caption{Summary of the properties of NGC\,1365--X1. The luminosity 
$L_{\rm x}$ is given for the 0.1--2.4 keV band.}  
     \label{fitres2}
      \begin{tabular}{cccc} 
      \hline
      \noalign{\smallskip}
\multicolumn{4}{l}{HRI position (J\,2000)}  \\
\multicolumn{4}{l}{~~~ $\alpha = 3^h 33^m 34\fs5,~~ \delta = -36\degr 
9\arcmin 38\farcs0$} \\
\multicolumn{4}{l}{spectral properties} \\
\multicolumn{4}{l}{~~~ \G $\simeq -1.5$ for $N_{\rm Gal}$, $L_{\rm x} \simeq 
2.4\,10^{39}$ erg/s (PSPC low-state) } \\
\multicolumn{4}{l}{variability}\\
\multicolumn{4}{l}{~~~ amplitude factor $\approxgt$ 10, within 
 $t$ $\approxlt$ 6 months}\\
      \noalign{\smallskip}
      \hline
 \end{tabular}
\end{table}

\section{Discussion}

\subsection{The core source} 

\subsubsection{Soft X-ray spectrum -- spectral fits}

A description of the soft X-ray spectrum in terms of a 
{\em single} PL (Sect.\,2.1, see also Fig. 1) 
leaves systematic residuals suggesting the 
presence of a 
thermal component. Unless metal abundances are strongly reduced (a factor 
$\sim 
5$ relative to
cosmic) a {\em single} RS model is unfeasible as well. All evidence from 
HII 
regions in NGC\,1365
(Alloin et al.\ 1981, Roy \& Walsh 1988, Zaritsky et al.\ 1994, Roy\& 
Walsh 1997) leads to 
an O/H and N/H of above-solar or close-to-solar  values
in the center and a slight decrease outwards.
The shallowness of the gradient is explained by
bar-induced
{\em strong mixing} of the gas.
Hence, 
strongly depleted metal abundances in the central X-ray gas are 
unexpected so that the successful metal-reduced single-RS fit is 
considered as an unlikely solution.{\footnote{For a recent discussion 
of the issue of single-component X-ray spectral models
of very subsolar abundances vs. two-component models of $\sim$solar abundances
see also Buote \& Fabian (1998).}}  

In addition, the requirement of at least two components is in agreement
with the \asca data analysis by I97.
Beyond the \ros band they found a significant hard component which shows 
up as a very hot RS component ($kT = 6.3$ keV)
in our RS+RS fit or is consistent with either a canonical
PL ($\Gamma_{\rm x}=-1.9$) plus a hot RS (1 keV) or a flat 
($\Gamma_{\rm x}=-1.0$)
PL and a `moderate' RS temperature (0.7 keV) in our RS+PL fits.
A warm-absorber model 
seen {\em purely} in absorption did not fit the spectrum,
either
suggesting the lack of highly ionized gas 
directly along the l.o.s 
or a very high column density.

Aside from the amount of cold absorption, the two-component fits and the
AGN warm-scatterer model (Sect.\ 2.1) are strikingly similar 
to our results for the ultraluminous
IR galaxy NGC\,6240 (Schulz et al.\ 1998, Komossa et al.\ 1998)
in which we proposed an AGN contribution in soft X rays.
For NGC\,1365, however, one of the arguments, the huge luminosity, cannot 
be used against
generation of the total X-ray flux in a starburst induced scenario.
Our RS+PL$-1.9$ model fit 
of the core source 
yields $L_{\rm 0.1-2.4 keV} = 6.1\,10^{40}$ erg/s 
($4.6\,10^{40}$ erg/s) for $N_{\rm H}$ free (fixed to 
$N_{\rm Gal}$). 
This is at least 
twenty (more likely more than forty) times less luminous 
than the central source of NGC\,6240.

In the following, to ease the luminosity demands, we use 
the solution for $N_{\rm H} = N_{\rm Gal}$ of which the
RS contribution is $2.4\,10^{40}$ erg/s or 52\% of the total.
The simplest idea one can have 
is to attribute the thermal RS component to
a starburst source and the PL to the AGN which means that they contribute 
in
equal proportions in the \ros band. We take these numbers as a
working hypothesis clearly having in mind that a precise decomposition
is not possible at the present stage of measurement resolution and theory.

\subsubsection{Clues from IRAS color indices} 
Are we able to constrain the relative amounts of the AGN and starburst 
contribution via multi-wavelength properties of NGC\,1365? Firstly, we 
formed
IR gradients 
$\alpha(\lambda_1,\lambda_2) = (\log f_{\nu_1} - \log f_{\nu_2})/(\log 
{\nu_1} - \log {\nu_2})$
from the IRAS flux densities $f_{\nu}$ (labeled by wavelength: 
$f_{25}=13.07$ Jy, $f_{60}=84.2$ Jy, $f_{100}=185.4$ Jy
($\sim \pm15\%$; retrieved from NED) at
$\lambda_i= 25, 60$ and 100$\mu$m and checked the location of NGC\,1365 
in 
the $\alpha(100,60)-\alpha(60,25)$ diagram. Samples of Seyfert class-1, 
class-2 and
HII-type objects in this diagram are shown in Miley et al.\,(1985) and 
Mulchaey et al.\,(1994). 
With $\alpha(100,60)=-1.54$ and $\alpha(60,25)=-2.13$, NGC\,1365 lies in a
region mainly occupied by HII-galaxies and Seyfert-2 galaxies.
The slope indices signify
dominating flux at smaller wavelengths (`warm' dust components), 
warmer than typical Seyfert-1s in the mid-IR (25,60)
and warmer than the {\em average} of HII region type galaxies in the far 
IR 
(100,60), making the FIR cool dust component (that dominates
in normal spirals) insignificant.
Altogether this suggests that the IR emission
comes from the central region where the AGN and relatively young star 
formation
components contribute a substantial portion of
the observed IR emission in NGC\,1365.

Its more X-ray and IR-luminous sister NGC\,6240 
contains a much
stronger cool-dust FIR-component ($\alpha(100,60)=-0.40$)
 but has nearly the same gradient in the
mid-IR ($\alpha(60,25)=-2.16$).

Putting NGC\,1365 with
$f_{60}/f_{25} = 6.44$ into the scheme of Heisler et al.\ (1997) shows a
discrepancy.
This scheme assumes that objects with $f_{60}/f_{25} > 4$ (cold mid-IR,
more edge-on view
of torus) should not even show a `hidden' BLR. But NGC\,1365 clearly exhibits broad
Balmer lines.
Converting the ratio into a power-law index $\alpha(60,25)$ (4
corresponds to $-1.6$) and checking
Mulchaey's et al.\ (1994) plot of $\alpha(60,25)$ versus $\alpha(100,60)$
 shows that at least
six broad-line objects lie in Heisler's et al.\ (1997) regime, where the
edge-on view on the torus
should hide the BLR. 

\subsubsection{Starburst contribution, superwind estimate} 

A 100\degr~ wide [OIII] enhanced region with 
line-splitting to the SE of the nucleus 
(Phillips et al.\ 1983, J{\"o}rs{\"a}ter \& Lindblad 1989) 
was kinematically modelled as an outflow cone
by Hjelm \& Lindblad (1996). This could be a Seyfert outflow, driven
by a wind from the active core, or a starburst outflow, driven by
a series of supernova explosions. Assuming that the outflow region
 evolved from a wind-driven
supershell of swept-up ISM we can use the MacLow \& McCray (1988) 
analytical model (as 
e.g. applied by Heckman et al.\ (1996) to Arp\,220 and 
Schulz et al.\ (1998) to 
NGC\,6240) for relating $L_{\rm 0.1-2.4 keV}$ 
(generated in the shell and the
bubble interior to it) to the mechanical input power $L_{\rm mech}$. If
starburst driven, this can be derived from the SN rate by scaling the 
power
of one SN per yr, $L_{\rm mech}= 10^{51}{\rm erg/yr}= 3.17\,10^{43}$ 
erg/s. 
Kristen et al.\ (1997) estimate $\sim 10^{-3}$ SN/yr for each of the two 
dominating brightest starburst knots 7\arcsec~ to the SW of the nucleus. 
The nucleus has the same $B$-band brightness as one of these knots and may
contain an equally strong mini burst plus the AGN.

As will be shown in Sect. 5.1.4, the optically detected AGN could only 
provide
a few percent of the IR. 
Assuming that starformation takes care of the rest  
we take the IRAS $L_{\rm IR}= 2.4\,10^{44}$ erg/s as an estimate
for the bolometric luminosity of the starburst for which
models from Gehrz et al.\ (1983)  predict a SN rate between 0.01
and 1 yr$^{-1}$ (depending on the IMF and the lower and upper mass limits)
and $L({\rm H}\alpha)$ in the range $6.9\,10^{41} - 6.9\,10^{42}$ erg/s
of which Kristen et al.\ see only $6\,10^{40}$ erg/s in the central 
region.
Hence, 90\% to 99\% of the H$\alpha$ emitters ionized by young stars are 
hidden.
Could this hidden burst supply the X-rays in a wind driven shell?
From the velocity field in Hjelm \& Lindblad (1996) we read off
a radius $R \sim 1$ kpc with a corresponding velocity 
${\rm d}R/{\rm d}t \sim 100$ km/s. With the thermal RS X-ray luminosity
of $2.4\,10^{40}$ erg/s we get the following solution of the
MacLow-McCray equations given in scaled form 
as in Schulz et al.\ (1998):  
\begin{equation}
R = (1\, \mbox{\rm kpc})\, L_{4.5E41}^{1/5}\, 
n_{1.2}^{-1/5}\, t_{6E6}^{3/5}~~,
\end{equation}
\begin{equation}
\frac{{\rm d}R}{{\rm d}t} = (100\,\mbox{\rm km s}^{-1}) 
\,L_{4.5E41}^{1/5}\, 
n_{1.2}^{-1/5}\, t_{6E6}^{-2/5}~~,
\end{equation}
\begin{equation}
L_{\rm 0.1-2.4 keV} = (2.4\,10^{40} \mbox{\rm erg s}^{-1})\, 
L_{4.5E41}^{33/35}\, 
n_{1.2}^{17/35}\,t_{6E6}^{19/35}~~.
\end{equation}
Here, $L_{4.5E41}$ corresponds to $L_{\rm mech}= 4.5\,10^{41}$ erg/s
or a SN rate of 0.015 yr$^{-1}$ which is consistent with the
Gehrz et al. starburst models. The expansion time $t_{6E6}$ corresponding
to $6\,10^6$ yrs and the density $n_{1.2}$ (1.2 cm$^{-3}$) are reasonable
for the central region of this strong bar. 

We conclude that this crude scenario can account 
for the observed thermal X-ray luminosity if more than 90\% of the ionized
gas emitting optical lines in the starburst region is obscured.
Due to less clear constraints an analogous estimate for an AGN wind is
not attempted.

\subsubsection{AGN contribution, IR-optical-X relations} 
We now compare the data with the sample of hard X-ray detected AGN 
from Piccinotti et al.\ (1982). Concerning the two excellent AGN tracers
broad lines and hard X-rays this sample is homogeneous and comprises 
QSOs, 
Seyfert-1 galaxies and
broad-line radio galaxies so that contributions by starburst components 
are
minimized. For the Piccinotti sample, Ward et al. (1988) 
showed that the luminosities
(determined for $H_0 = 50$ km/s/Mpc)    
of hard X-rays,
H$\alpha$ and  mid-IR-radiation\footnote{Ward et al. (1988) derived the
IR ``luminosity'' $L(25-60\mu {\rm m})$ from the flux parameter 
$f_{25-60\mu{\rm m}} \equiv \nu_{25} f_{25} + \nu_{60} f_{60}$}   
are well correlated (their Figs. 5 and 6).
We represent their results by the relations
\begin{eqnarray} 
\log L_{\rm 2-10 keV} & = & 0.947 \log L({\rm H}\alpha) + 3.447~~, \nonumber 
\\
\log L_{\rm 2-10 keV} & = & 1.326 \log L(25-60\mu {\rm m}) - 15.283~~.
\end{eqnarray}
The 
scatter in the ordinate is $\pm0.3$ for both relations.

For NGC\,1365 these quantities are observed to be (cgs-quantities scaled 
to $H_0$=50 for the
comparison):
$\log L_{\rm 2-10 keV} = 41.08$ (from I97);
$\log L({\rm H}\alpha) = 40.82$ (broad component measured by Schulz et 
al.\ 1994);
$\log L(25-60\mu {\rm m}) = 44.87$ (derived from IRAS data given by NED).
It turns out that these combinations of data do not fit the 
relations 
of the
Piccinotti sample for which we assume that it defines
`pure' AGN properties. E.g., if the hard X-rays come from the AGN in 
NGC\,1365
we would expect $\log L(25-60\mu {\rm m}) = 42.50$ which is only 0.43\% 
of the
observed 44.87. The remaining IR could either be attributed to star 
formation
or the source for the hard 2-10 keV photons is for a large part obscured 
which would, however,
require column densities exceeding $10^{24}$ cm$^{-2}$.

Utilizing the first Piccinotti-sample relation of Eq.\ 4, 
the X-rays predict $\log L({\rm H}\alpha)=39.74$, only 8.3\% of the 
observed
40.82. The data were taken at different epochs, so variability cannot be 
excluded to explain part
of the discrepancy. However, the data for the Piccinotti sample were 
collected non-simultaneously as well 
and their total scatter corresponds to only a factor four. Another 
possibiIity
would be some arrangement of obscuring material in front of
the X-ray source or/and the BLR possibly combined with
scattering elsewhere. Compton thick and Balmer electron-scattering
highly ionized thick columns
might as well be envisioned in the extreme environment of the nucleus,
but detailed models are beyond the scope of this work.
Taking H$\alpha$ as representative
for the AGN, $\log L_{\rm 2-10 keV} = 42.10$, i.e.\ ten times the 
observed hard X-ray
luminosity, and $\log L(25-60\mu {\rm m}) = 43.28$, i.e.\ 2.5\% of the 
observed mid-IR,
would be predicted. This type of solution would fit to our
warm scatterer model (last model in Sect.\,2.2) which could explain both the 
hard PSPC
(and {\sl ASCA}) PL-like component and the high equivalent width of the 
FeK 
complex (like in NGC\,6240, see Komossa et al. 1998).

A crude upper limit for the contribution from a
stationary AGN would be one that dominates the mid-infrared, leading to 
the prediction
of $\log L$ = 44.22 and 43.05 for the hard X-ray and H$\alpha$ 
luminosities, 
respectively. In this case, the corresponding
observed (see above) hard X-rays and H$\alpha$ luminosities are 
0.06\% of the 
predicted ones 
requiring esentially complete obscuration of the emission-line and hard 
X-ray sources.
Such an extreme situation appears to be ruled out for NGC\,1365 because 
star formation is spatially extended over a diameter of $\sim 14\arcsec$
and will provide an appreciable part of the IR. 

\subsubsection{Place within the unified model} 

NGC\,1365
does not fulfill expectations of the simplest version of the unified model
in which only the torus blocks the light and detection of a BLR would
imply an unobscured view of the X-ray source as well.  
Complicated models appear to be necessary to let this AGN be an
{\em intrinsically normal} broad-line object.
Other, currently vague, possibilities would be
that the AGN in NGC\,1365 as a low-luminosity object is qualitatively
different from the Piccinotti objects or that this object is an example 
for non-mainstream scenarios, like the
starburst-AGN models developed by Terlevich et al.\ (1992). However, such 
models have not yet been worked
out to satisfactorily account for the hard X-ray data.

\subsection{The bright and highly variable off-nuclear 
source NGC\,1365--X1}
NGC\,1365--X1 is a bright X-ray source. 
The excellent spatial resolution of the
ROSAT HRI locates it on one of the spiral arms
of NGC\,1365.
The source is highly variable on the
timescale of months.  

In Sect. 4 we already gave cautious comments on the possibility 
of a foreground or background source and considered it unlikely.
In the following discussion, we assume the source to be  
intrinsic to NGC\,1365.  
In this case, NGC\,1365--X1 is one of the 
brightest off-nuclear point sources
known, with an isotropic luminosity of 
$L_{\rm 2-10 keV} \approxgt 4\,10^{40}$ erg/s (for $d = 19.8$ 
Mpc) in
the source's observed high-state. 

Off-nuclear X-ray sources, some of them variable, 
are frequently discovered in nearby galaxies
(e.g. Fabbiano 1989, Vogler 1997;  Turner et al. 1993, Ehle et al. 1995,
Brandt et al. 1996, Vogler et al. 1997, 
Supper et al. 1997, Immler et al. 1998).
Their X-ray luminosities
are typically of order 10$^{38}$ erg/s up to several 10$^{39}$ erg/s.
Among the most X-ray bright observed sources in other galaxies are 
supernovae 
(of type II; see e.g. the review by Schlegel 1995).
The reported maximum X-ray luminosities range
between several 10$^{39}$ erg/s (e.g. SN 1993J, Zimmermann et al. 1994b;
SN 1979C, Immler et al. 1998)
and 3\,10$^{40}$ erg/s (SN 1986J, Bregman \& Pildis 1992; all luminosities
are as given by the authors).   
On long timescales, variability is detected in some SN (cf. 
Schlegel 1995).
Within our Galaxy, one of the most luminous individual X-ray emitters is
the superluminal-motion source GRS 1915+105 with $L_{\rm x} \simeq 
10^{39}$ erg/s
(daily average during high-states) and peak luminosities
up to $5\,10^{39}$ erg/s (on a timescale of seconds;
Greiner et al. 1998).
  
NGC\,1365--X1 is exceptionally luminous.
Its temporal variability excludes an interpretation in terms of several
spatially unresolved weak sources.
Another definitely bright ($L_{\rm x} \approx 10^{40}$ erg/s)
source was found by Vogler et al. (1997) in an outer 
spiral arm of NGC 4559. They suggested an interpretation in terms of
a `buried supernova'. 
In contrast to NGC\,1365--X1, this source did not show any
X-ray variability.
Although a supernova in dense medium is an efficient way to reach 
high X-ray luminosities (e.g. Shull 1980, Wheeler et al. 1980,
Terlevich et al. 1992), the huge variability on the timescale
of months we detect seems to favour an interpretation in terms
of accretion onto a compact object. 
The amplitude and rapidity of variability share similarity
with the off-nuclear source in NGC\,4945 reported by 
Brandt et al. (1996). However, the present case is much more
powerful, by at least a factor of $\sim$16 in luminosity.{\footnote{They
observe $L_{\rm max} \simeq 8\,10^{38}$ erg/s in the 0.8--2.4 keV band,
NGC\,1365--X1 shows $L_{\rm max} \simeq 1.3\,10^{40}$ erg/s in the same band.}}   

At present, the most likely interpretation seems to be an ultra-powerful  
X-ray binary, with either a highly super-eddington low-mass black hole
or a very massive black hole.  
The hard powerlaw component derived from the \ros and \asca spectral
fits is known to be present in Galactic X-ray binaries (see Tanaka 1997
for a recent review);
they have soft-excesses added in their high-states. 
In case the accretion is not super-eddington, a rather
high-mass black hole of $\sim$100--200 M$_{\odot}$ is inferred
which may pose a 
challenge
for stellar evolution models.

\section {Summary and conclusions}

We studied the nuclear and circumnuclear X-ray emission of NGC\,1365
using \ros PSPC and HRI observations.  
The HRI image shows compact emission from the nucleus,
from the highly variable southwestern source, and resolves the source
just outside the SW spiral arm into a double source. 

The analysis of the spectral 
and temporal X-ray properties
of the nucleus of NGC\,1365 and the 
enigmatic source RX\,J0333-3609 = NGC\,1365--X1 is summarized as follows:\\

\noindent {\em The core of NGC\,1365.}
To describe the X-ray spectrum, we favour a two-component model consisting
of about equally strong contributions from a Raymond-Smith (RS) plasma 
(with $\sim$cosmic abundances) and a powerlaw. 
Although a single RS model yields a successful spectral fit as well, it
requires heavily depleted metal abundances which seems to be in contradiction 
to optical observations.  
The powerlaw component signifies the presence of
a hard component not common in simple starbursts; this
hard component as well as an FeK line is also present in
ASCA data of I97
strengthening the view that it
arises in the AGN. 

The thermal RS part of the X-ray spectrum is found to be consistent with
being produced in starburst-induced superwind shocks if the starburst
dominates
the mid to far IR and its H$\alpha$ emitters are largely hidden.

A detailed comparison of the X-ray properties
with multi-wavelength observations of NGC\,1365 
and a large sample of type-I AGNs
was performed.
We find that, if the hard X rays (2--10 keV) or the broad H$\alpha$ are representative 
tracers of the
active nucleus in NGC\,1365, they can only account for a minute amount 
of the IR; or, conversely, if the IR were due to dust heating by the AGN, the latter has 
to be
essentially completely hidden. This and the large extent of the 
starformation region support
the view that the IR is starburst dominated although emission-line 
H$\alpha$
from the starburst has to be largely obscured as well. 

A puzzle is the low X-ray luminosity of NGC\,1365
as compared to the strength of the BLR Balmer line component;
this poses a problem for the simplest version of the
unified model. 
In particular, a comparison with the 
Piccinotti sample of type-1 AGNs strengthens the view that we see
scattered X-rays in the 2--10 keV range.  This
is in line with our alternative successful X-ray spectral model of 
a warm scatterer; highly photoionized and high column density material close to
the AGN continuum source.  
Such a scenario, as alternative to a low-luminosity AGN viewed directly,
still requires a peculiar viewing geometry to simultaneously
explain the observed strength of the broad lines. \\

\noindent {\em NGC\,1365--X1.} Attributed to NGC\,1365, this source 
is among the most 
luminous ($L_{\rm x,max} \ge$ several 10$^{40}$ erg/s) 
and most highly variable (factor $\ge$ 10 within 6 months)   
off-nuclear X-ray sources known so far.
At present, the most likely interpretation seems to be an ultra-powerful
X-ray binary. 
If this interpretation is confirmed, 
this or similar X-ray sources in nearby
galaxies will be important probes for the extremes of stellar
evolution.
Further monitoring of NGC\,1365--X1, and derivation of improved 
spectral information  will certainly be worthwhile.

\begin{acknowledgements}
St.K. acknowledges support from the Verbundforschung under grant No. 50\,OR\,93065.
It is a pleasure to thank Per Olof Lindblad for helpful comments and suggestions. 
The \ros project is supported by the German Bundes\-mini\-ste\-rium
f\"ur Bildung, Wissenschaft, Forschung und Technologie 
(BMBF/DLR) and the Max-Planck-Society.
We thank Gary Ferland for supplying {\em Cloudy}, and Andreas Vogler
for providing the software to plot the overlay contours of Fig. 4.
The optical image shown is based on photographic data of the 
National Geographic Society -- Palomar
Observatory Sky Survey (NGS-POSS). 
This research has made use of the NASA/IPAC extragalactic database (NED)
which is operated by the Jet Propulsion Laboratory, Caltech,
under contract with the National Aeronautics and Space
Administration.
\end{acknowledgements}

\end{document}